# Interictal MEG abnormalities to guide intracranial electrode implantation and predict surgical outcome

Tom Owen[1], Vytene Janiukstyte[1], Gerard R. Hall[1],
Fahmida A. Chowdhury[3,5], Beate Diehl[3,5], Andrew McEvoy[3,5],
Anna Miserocchi[3,5], Jane de Tisi[3,4,5], John S. Duncan[3,4,5],
Fergus Rugg-Gunn[3,5], Yujiang Wang[1,2,3,5], Peter Neal Taylor[1,2,3,5]

1. CNNP Lab (www.cnnp-lab.com), Interdisciplinary Computing and Complex BioSystems Group, School of Computing, Newcastle University, Newcastle upon Tyne, United Kingdom

2. Faculty of Medical Sciences, Newcastle University, Newcastle upon Tyne, United Kingdom

3. UCL Queen Square Institute of Neurology, Queen Square, London, WC1N 3BG, United Kingdom

4. NIHR University College London Hospitals Biomedical Research Centre, UCL Queen Square Institute of Neurology, London WC1N 3BG, United Kingdom

5. National Hospital for Neurology & Neurosurgery, Queen Square, London, WC1N 3BG, United Kingdom

* t.w.owen1@newcastle.ac.uk  & peter.taylor@newcastle.ac.uk

# Abstract

Intracranial EEG (iEEG) is the gold standard technique for epileptogenic zone (EZ) localisation, but requires a preconceived hypothesis of the location of the epileptogenic tissue. This placement is guided by qualitative interpretations of seizure semiology, MRI, EEG and other imaging modalities, such as magnetoencephalography (MEG). Quantitative abnormality mapping using MEG has recently been shown to have potential clinical value. We hypothesised that if quantifiable MEG abnormalities were sampled by iEEG, then patients' post-resection seizure outcome may be better.

Thirty-two individuals with refractory neocortical epilepsy underwent MEG and subsequent iEEG recordings as part of pre-surgical evaluation. Eyes-closed resting-state interictal MEG band power abnormality maps were derived from 70 healthy controls as a normative baseline. MEG abnormality maps were compared to iEEG electrode implantation, with the spatial overlap of iEEG electrode placement and cerebral MEG abnormalities recorded. Finally, we assessed if the implantation of electrodes in abnormal tissue, and subsequent resection of the strongest abnormalities determined by MEG and iEEG corresponded to surgical success.

Intracranial electrodes were implanted in brain tissue with the most abnormal MEG findings - in individuals that were seizure-free post-operatively (T=3.9, p=0.003), but not in those who did not become seizure free. The overlap between MEG abnormalities and electrode placement distinguished surgical outcome groups moderately well (AUC=0.68). In isolation, the resection of the strongest abnormalities as defined by MEG and iEEG separated surgical outcome groups well, AUC=0.71, AUC=0.74 respectively. A model incorporating all three features separated surgical outcome groups best (AUC=0.80).

Intracranial EEG is a key tool to delineate the EZ and help render individuals seizure-free post-operatively. We showed that data-driven abnormality maps derived from resting-state MEG recordings demonstrate clinical value and may help guide electrode placement in individuals with neocortical epilepsy. Additionally, our predictive model of post-operative seizure-freedom, which leverages both MEG and iEEG recordings, could aid patient counselling of expected outcome.

# Introduction

Intracranial EEG (iEEG) recordings are widely considered as the gold-standard technique to accurately localise the epileptogenic zone (EZ - the part of the brain indispensable for seizures). Multiple markers of the EZ have been developed from interictal spikes[1–4] and high frequency oscillations[5–12], to the ictal onset patterns themselves[13]. Successful iEEG implantation requires a preconceived hypotheses of the location of epileptogenic tissue. Thus, if the EZ is not implanted one may expect poorer post-surgical outcomes.

The planning of iEEG electrodes depends on seizure semiology, MRI, scalp EEG, and MEG. Magnetoencephalography (MEG) recordings may aid electrode implantation; however, most analyses largely remain qualitative and mainly investigating spikes[14–19]. Band power abnormality maps from interictal MEG data recently quantified epileptogenic tissue in individuals with refractory neocortical epilepsy using a data-driven framework, and demonstrated localisation overlap with subsequent resection only in seizure-free patients[20]. With complete cortical coverage, and sensitivity to abnormalities, MEG band power abnormality maps may be of use to localise the EZ and guide intracranial electrode placement.

Although both modalities capture neurophysiological activity, iEEG and MEG are differentially sensitive to sources of activity, and thus provide complementary information. As pyramidal cells are organised perpendicular to the cortex, iEEG typically reflects extracellular sources whilst MEG reflects intracellular sources[21]. As such, iEEG and MEG are more sensitive to sources located at the crowns of gyri, and sulci and fissures respectively, depending on placement[22]. It is possible that multimodal abnormality mapping may provide a more complete view of the epileptogenic zone and thus further aid clinical decision making.

In this study we performed a multimodal analysis to investigate two primary hypotheses. First, we quantified if intracranial electrodes were implanted in regions of high MEG abnormality, and hypothesised a greater overlap in patients who were seizure-free after resection. Second, we hypothesised that if electrodes were implanted in regions of high abnormality, then seizure-freedom would be expected if the greatest abnormalities in both modalities were also resected.

# Methods

## Patient and control data

We retrospectively analysed data from 32 individuals with refractory neocortical epilepsy who underwent resective surgery. Surgical success was defined using the International League Against Epilepsy (ILAE) scoring[23] one year post-operatively. Twelve individuals were entirely seizure-free after surgical intervention (ILAE 1). No significant differences were present between surgical outcome groups based on age, sex and epilepsy duration (Table 1). All individuals underwent pre-operative MEG and then subsequent intracranial EEG recordings as part of their pre-surgical evaluation. Additionally, T1-weighted MRI scans were acquired for each individual both pre and post-operatively. For normative baselines, 70 healthy controls underwent eyes-closed resting-state MEG recordings in Cardiff[24] and 234 individuals underwent invasive intracranial recording as part of the RAM dataset.

**Table 1: Summary of clinical demographics by surgical outcome groups.** The Mean and standard deviations are reported, Mean(SD), for seizure-free (ILAE 1) and non seizure-free (ILAE 2+) individuals. Statistical tests were performed to assess whether any differences exist between the groups. For continuous variables, two-tailed t-tests were used. For categorical features, two-tailed Chi-squared tests were used.

|  | Seizure-free (ILAE1) | Not seizure-free (ILAE2+) | Significance |
|---|---|---|---|
| N | 12 | 20 |  |
| Age (years) | 30.5 (7.0) | 32.3 (11.3) | p=0.636 |
| Sex (Female/Male) | 3/9 | 10/10 | p=0.895 |
| Epilepsy duration | 20.5 (8.2) | 20.0 (8.8) | p=0.861 |

## MRI preprocessing

Pre and post-operative MRI scans were acquired for each subject with refractory epilepsy and were used to delineate their resections. In short, MRI scans were acquired using a 3T GE Signa HDx scanner using standard imaging gradients, a maximum strength of $40\text{mT}m^{-1}$ and slew rate of $150\text{T}m^{-1}s^{-1}$. Data were acquired using a body coil for transmission, and an 8 channel phased array coil for reception. Standard clinical sequences were performed including a coronal T1-weighted volumetric acquisition with 170 contiguous 1.1 mm-thick slices (matrix, 256 × 256; inplane resolution, 0.9375 × 0.9375 mm). Individual MRI scans were preprocessed using Freesurfer's pipeline 'recon-all'[25] and subsequently parcellated into 114 neocortical regions of interest (ROI) based on the Lausanne parcellation scheme[26]. To delineate the resection cavity, pre- and post-operative MRI scans were linearly co-registered using FSL and overlaid[27–29]. Resection volumes were manually drawn for each individual using FSLview and pre and post-operative volumes were estimated using custom MATLAB code[30]. A region was defined as resected if the pre- and post-operative volume change exceeded 10%. Regions with a volume change less than 1% were defined as spared, with remaining regions (volume change 1-10%) excluded from the analysis. Healthy individuals at Cardiff also underwent T1-weighed MRI acquisition using a 3T GE Signa HDx scanner. A full description of the acquisition protocol has been described previously[24].

## MEG processing and abnormality mapping

MEG recordings for patients and healthy control cohorts were acquired using a 275 channel CTF whole head MEG system at different sites. Resting-state eyes-closed interictal recordings for subjects with epilepsy were acquired at UCL in London, and for healthy control data at CUBRIC Cardiff as part of the MEG UK partnership. MEG recordings from both cohorts were processed in Brainstorm using previously described methods[20]. MEG sensor locations were co-registered to the individuals' MRI scan using fiducial points. Following co-registration, MEG recordings were downsampled to 600Hz and cleaned of any arifacts. Powerline artifacts were removed between 47.5-52.5Hz using a notch filter, and ocular and cardiac artifacts were removed manually using independent component analysis (ICA). Once cleaned of any arifactual noise, MEG recordings were source reconstructed using the minimum-norm imagaing technique, sLORETA[31], and an overlapping spheres headmodel. Subsequent source space time-series were downsampled into cortical regions of interest (ROIs) using the Lausanne parcellation scheme[26]. Finally, 70 second epochs of recordings clear of residual artifacts for each individual were used to construct neocrotical maps of band power abnormalities.

To construct normative maps, regional power spectral densities were computed using Welch's method using a two second sliding window with one second overlap. Regional relative band power estimates for delta (1-4Hz), theta (4-8Hz), alpha (8-13Hz), beta (13-30Hz), and gamma (30-80Hz, excluding 47.5-52.5Hz) were averaged across all seventy healthy controls (Figure 1B). Individual band power abnormality maps were derived for each region by computing the absolute z-score relative to normative baselines for each of the five frequency bands. To reduce the dimensionality of the data we retain the maximum regional absolute z-score across frequency bands (Figure 1D).

## iEEG processing and abnormality mapping

Long-term iEEG recordings were acquired for each individual prior to resective surgery. A cohort of 234 subjects acquired as part of the RAM dataset were used to construct the normative map, using contact recordings from outside of the seizure onset and propagation zone. Seventy second epochs of resting-state wakeful recordings were chosen for each individual. Similar to MEG (Section 2.3), relative band power contributions for each contact were computed for delta (1-4Hz), theta (4-8Hz), alpha (8-13Hz), beta (13-

30Hz), and gamma (30-80Hz excluding 47.5-52.5Hz and 57.5-62.5Hz). Note that artifacts were removed to account for interference from both US and UK powerlines (Figure 1C).

Intracranial electrodes were localised to ROIs using standard procedures [32]. In short, electrodes used to construct the normative map were converted from the Talairach coordinate system to standard MNI space and assigned to an ROI in the Lausanne parcellation based on the minimum euclidean distance. For the patient cohort, electrode assignment was performed in native space using pre-operative CT and MRI scans[33]. Similar to MEG, electrodes were considered resected if the pre and post-operative volume change of the region exceeded 10%.

## Overlap between MEG abnormalities and electrode placement

We hypotheised that intracranial electrodes were implanted in regions with the greatest MEG abnormality in individuals who were seizure-free post-operatively. To quantify the degree of overlap between MEG abnormalities and intracranial electrode placement we used the abnormality coverage. Similar to the $D_{RS}$[33], the abnormality coverage quantifies the degree in which electrodes are placed in tissue of strongest MEG abnormality using the area under the receiver-operating characteristic (ROC) curve (AUC). Ranging between zero and one, an abnormality coverage of 1 corresponds to electrodes implanted exclusively in the most abnormal neocortical tissue. Conversely, an abnormality coverage of 0 corresponds to the electrode implantation targeting the least abnormal neocortical tissue. An abnormality coverage of 0.5 corresponds to chance and reflects the targeting of both abnormal and seemingly healthy tissue (Figure 1E).

## Overlap between neurophysiological abnormalities and resection masks

To assess whether the locations with greatest abnormalities were resected we used the $D_{RS}$[20,32]. Like the abnormality coverage, the $D_{RS}$ ranges from zero to one, with values of zero corresponding to the resection of the most abnormal tissue. The $D_{RS}$ was computed for each individual using MEG and iEEG data separately using only tissue where there was MEG and electrode coverage i.e. discarding neocortical tissue in MEG where electrodes were not implanted (Figure 1F).

## Modelling of post-operative seizure freedom

We investigated the extent to which the abnormality coverage and two $D_{RS}$ measures explain surgical outcome using a logistic regression model. No standardisation in the form of mean centring and scaling was performed prior to model fitting as all three features have natural interpretations and similar values ranges. Class weights were introduced to the model in order to account for the imbalance in surgical outcome groups (12 ILAE 1 and 20 ILAE 2+). Setting a class weight of $\frac{20}{32}$ and $\frac{12}{32}$ for seizure-free and non-seizure-free groups respectively penalises the most frequent surgical outcome group (ILAE 2+) in such a way that both groups are treated equally. We report the output of the model using a nomogram (Figure 1G). In the context of epilepsy nomograms have previously been proposed to aid clinicians determine post-surgical seizure-freedom[34,35] and cognitive decline[36,37]. Nomograms are commonly used as a visual representation of the Cox proportional hazard model used in survival analysis; however, they can also be used for a logistic regression model. For a given subject, each feature within the nomogram accrues points towards a final score. The number of points attributed to each feature is directly proportional to the feature importance estimated from the logistic regression model. Once all of the points for a given subject are totalled, a prediction of surgical outcome can be made based on whether the patient exceeds a given threshold determined during model training. For the nomogram presented in this study, the more points a subject accrues, the greater the confidence that they will be seizure-free post-operatively.

To assess the robustness of the predictive model to outliers in the data we used leave one out validation. During leave one out validation, a single subject is removed from the dataset, the model is recomputed, and the AUC estimated. Once complete, the AUC scores are then averaged to obtain a robust measure of the separability of surgical outcome groups.

## Statistical testing

Statistical tests were used to assess whether the abnormality coverage and $D_{RS}$ scores differ significantly from chance. We used a one-tailed t-test to check whether the abnormality coverage of seizure-free patients was significantly greater than 0.5. A one-tailed Mann-Whitney U test was used to quantify whether our measures significantly separated surgical outcome groups. One-tailed tests were used as clear hypotheses of direction are provided.

## Code and data availability

Data and code to reproduce the figures is available upon reasonable request.

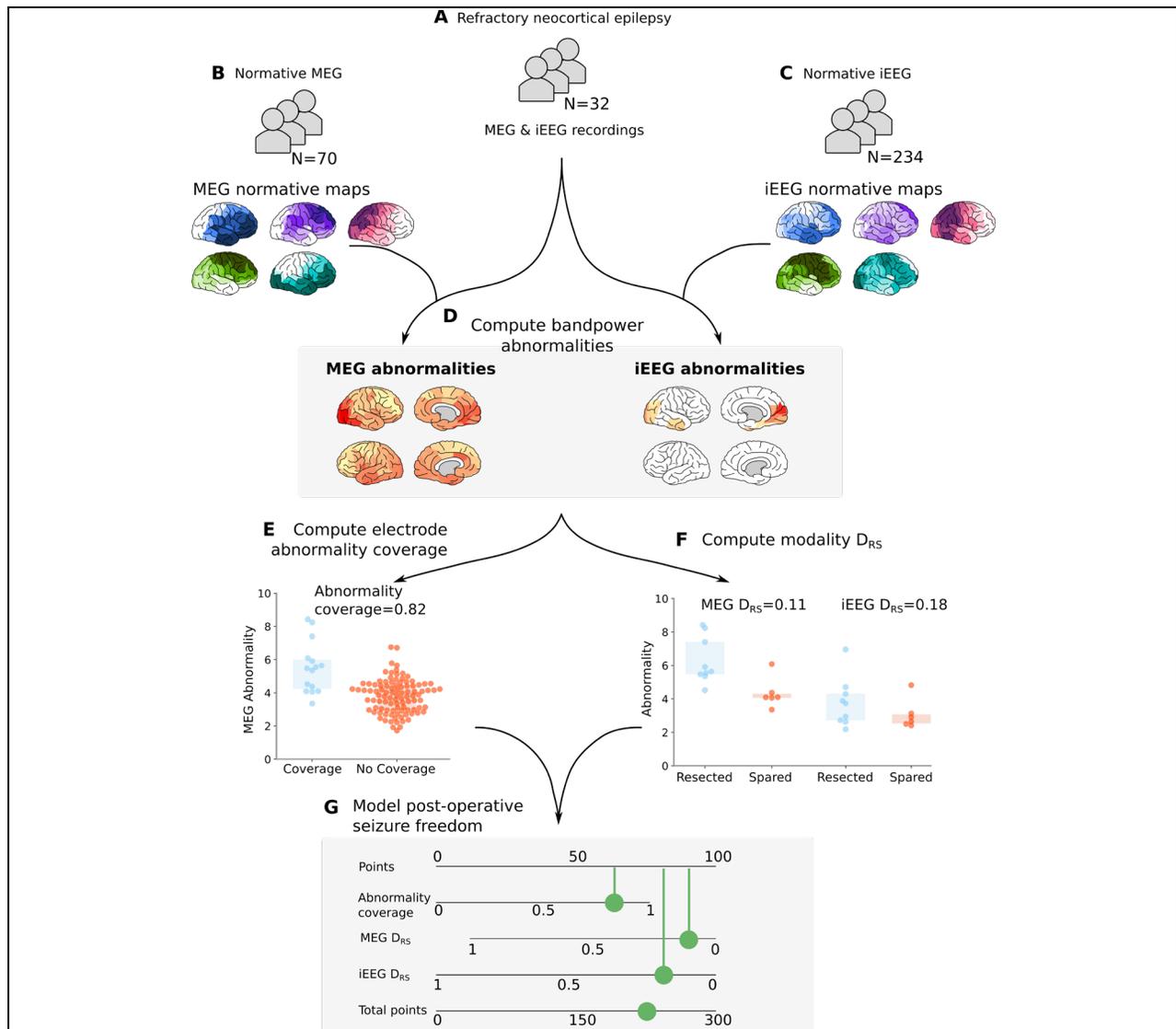

*Figure 1: Processing pipeline to assess the clinical utility of MEG band power abnormalities to guide iEEG implantation. (A-C) MEG and iEEG recordings were collected for healthy and patient cohorts. Recordings for 70 healthy controls and 234 individuals with epilepsy were used as a normative baselines for MEG and iEEG respectively. MEG and iEEG recordings were collected for an independent cohort of 32 individuals with refractory neocortical epilepsy. Regional relative band power was averaged across individuals and frequency bands to create normative maps. Patient maps of band power were derived using normative data as baselines by retaining the maximum absolute z-score across frequencies within each region (D). The overlap between the strongest MEG abnormalities and electrode placement was quantified, defined as the abnormality coverage, with values closer to 1 corresponding to the implantation in the most abnormal tissue (E). The resection of the strongest abnormalities defined by MEG and iEEG were quantified using the $D_{RS}$ (F). $D_{RS}$ values closer to 0 correspond to the resection of the strongest abnormalities. The $D_{RS}$ was only computed using neocortical tissue with MEG and iEEG coverage. The abnormality coverage and $D_{RS}$ values per individual were used to classify post-operative seizure-freedom using a logistic regression model. Model output is visualised using a nomogram (G), with each measure accruing points depending on the feature weight. The more points a subject accrues the more likely they are to be classified as seizure-free.*

# Results

## MEG abnormalities overlap with electrode placement in seizure-free patients

We investigated whether intracranial EEG electrodes were implanted in areas of strongest MEG abnormality using the 'abnormality coverage' metric. Two example subjects are shown in Figure 2 with different surgical outcomes. In the seizure-free patient (Figure 2A) a strong overlap exists between the iEEG electrode implantation and strongest MEG abnormalities. This is quantified with an abnormality coverage score of 0.82, signifying that electrodes were indeed implanted in the most abnormal neocortical tissue, as defined by resting-state interictal MEG band power.

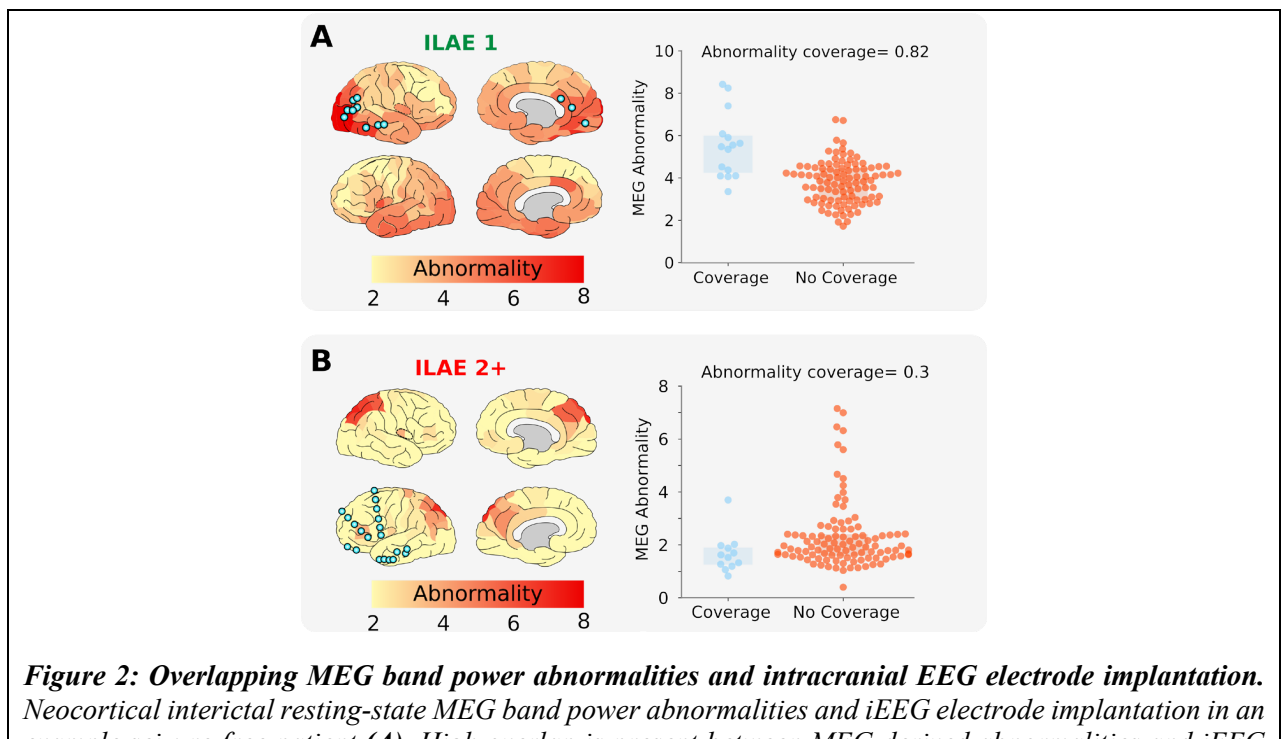

*Figure 2: Overlapping MEG band power abnormalities and intracranial EEG electrode implantation. Neocortical interictal resting-state MEG band power abnormalities and iEEG electrode implantation in an example seizure-free patient (A). High overlap is present between MEG derived abnormalities and iEEG electrode placement, quantified with an abnormality coverage of 0.82. In this scenario we would expect post-operative seizure freedom as iEEG electrodes have targeted abnormal tissue presumed to contain the epileptogenic zone. (B) Conversely, this example subject with poor surgical outcome (ILAE 2+) has minimal overlap between MEG abnormalities and electrode placement (abnormality coverage=0.3). As such, we would expect poor surgical outcome as the presumed epileptogenic tissue was not targeted by intracranial electrodes for further monitoring. Spatial heatmaps correspond to MEG derived band power abnormalities, with blue points corresponding to the approximate localisation of iEEG electrodes. Boxplots (right panels) illustrate the abnormality of regions with, and without iEEG coverage (blue and orange respectively). Each data point corresponds to a single neocortical region of interest. The abnormality coverage (0.82 for patient A) reflects if the most abnormal regions had iEEG coverage. Values closer to 1 corresponding to implantation exclusively in the most abnormal tissue and values of 0 to an implantation exclusively in the least abnormal tissue.*

In contrast to the seizure-free individual, figure 2B illustrates the overlap between iEEG electrodes and MEG abnormalities for a non-seizure-free subject. It is clear that electrode implantation does not overlap well with the MEG derived abnormalities, with the strongest abnormalities located in right occipital and parietal tissue, and electrodes implanted in left frontal tissue. The minimal overlap between iEEG electrode placement and MEG band power abnormalities is captured by the abnormality coverage measure with a value of 0.3.

We expanded our analysis to the full cohort of 32 individuals, reporting the overlap between electrode placement and MEG abnormalities (Figure 3). Individuals who were seizure-free post-operatively had greater overlap between MEG band power abnormalities and electrode placement, characterised by larger abnormality coverage values, than non-seizure-free individuals. The implantation of electrodes in tissue of strongest MEG abnormality occurred in seizure-free patients (ILAE 1) greater than chance (T=3.9, p=0.003). The effect of electrodes overlapping with MEG abnormalities separates surgical outcome groups well (AUC=0.68).

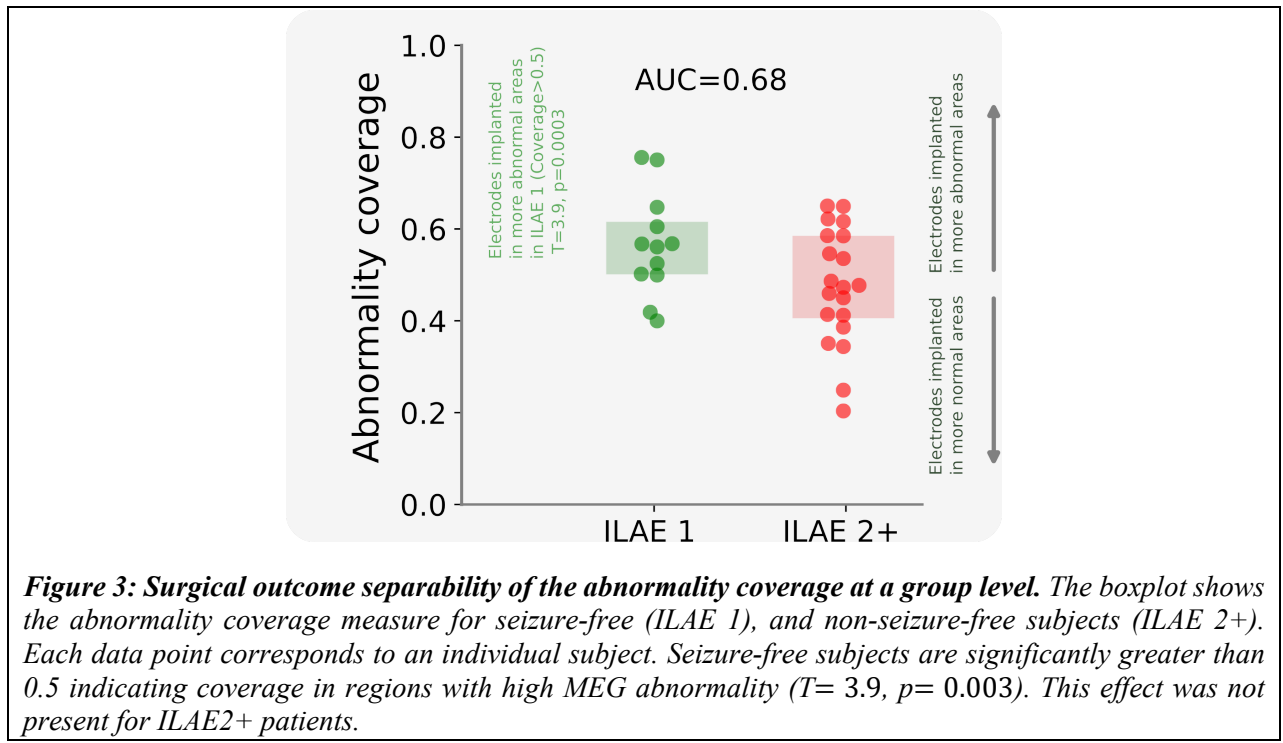

*Figure 3: Surgical outcome separability of the abnormality coverage at a group level. The boxplot shows the abnormality coverage measure for seizure-free (ILAE 1), and non-seizure-free subjects (ILAE 2+). Each data point corresponds to an individual subject. Seizure-free subjects are significantly greater than 0.5 indicating coverage in regions with high MEG abnormality (T= 3.9, p= 0.003). This effect was not present for ILAE2+ patients.*

Taken together, these results suggest that patients had better outcomes if their MEG-derived abnormalities were sampled by intracranial EEG.

## Multimodal abnormality maps predict post-operative seizure freedom

We next investigated if the strongest MEG and iEEG abnormalities were resected. To quantify this we used the $D_{RS}$ metric, considering only tissue which had MEG and iEEG coverage (Figure S1). In agreement with our prior studies[20,32], the resection of the strongest abnormalities were typically observed in seizure-free patients. The effect separating surgical outcome groups well for both MEG, AUC=0.71, and iEEG, AUC=0.74. Subject data and measures are reported in Supplementary Table 1.

We hypothesised that the combination of abnormality coverage and two $D_{RS}$ measures would explain surgical outcome best. Our rationale was that $D_{RS}$ would perform best for seizure-free patients only if abnormalites were actually covered, hence the inclusion of all three metrics. To combine measures, we used a logistic regression model and report the output as a nomogram (Figure 4A). All three measures contributed towards the prediction of post-operative seizure-freedom. The implantation of electrodes in MEG-defined abnormal regions, and subsequent concordance between MEG and iEEG, separated outcome groups best (Figure 4B). Robust measures of model performance using a leave one out approach resulted in an average AUC of 0.79 [Min=0.77, Max=0.84].

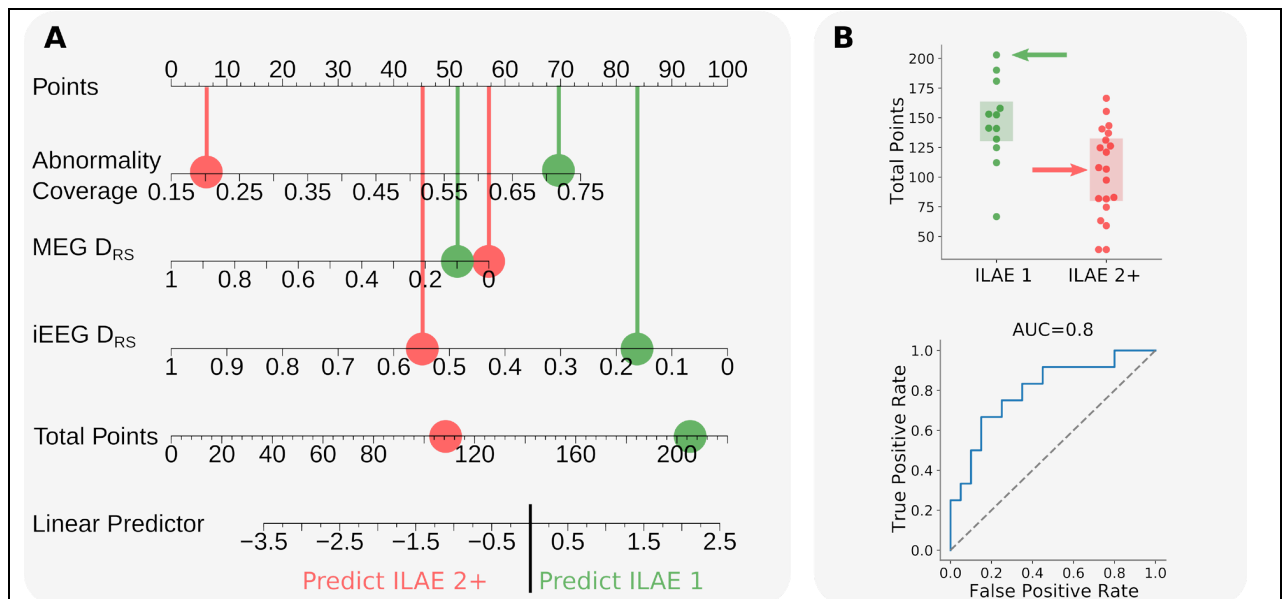

*Figure 4: Modelling post-surgical seizure-freedom using multimodal measures. (A) Nomogram illustrating the output of a logistic regression model trained using the abnormality coverage, MEG $D_{RS}$, and iEEG $D_{RS}$. Each feature accrues points towards a final score. The points for an individual subject based on their measures are totalled and subsequently compared across surgical outcome groups. Each green point corresponds to the results for a single seizure-free patient, whereas red points correspond to the results for a single non-seizure-free subjects. We hypothesised that the more points a subject accrued, the more likely they would be seizure-free post-operatively as the abnormality coverage indicates that potentially epileptogenic tissue had been targeted for iEEG monitoring and that MEG and iEEG are in agreement that that most abnormal tissue was resected. (B) For each individual, the total points calculated using the nomogram were compared across surgical outcome groups. The model results are presented as*

*a boxplot and receiver operating characteristic (ROC) curve. Each point corresponds to a single individual (ILAE 1: green, ILAE 2+: red).*

Together, the results of this study suggest that resting-state interictal MEG band power abnormality mapping may provide localising information which can be leveraged for electrode implantation. Furthermore, we show that a multimodal model (incorporating both iEEG and MEG) offers clinically interpretable predictions which may be of value during the pre-surgical evaluation.

## Discussion

Accurate delineation and resection of epileptogenic tissue is key to achieve post-operative seizure-freedom[38]. Intracranial EEG is widely used to delineate the EZ in difficult to localise individuals. Hypotheses of epileptogenic tissue location are required in order to guide electrode implantation. In this study we demonstrated that data-driven measures of neocortical abnormality using interictal MEG band power are associated with electrode implantation strategies in successful surgery candidates. Moreover, we showed that a multimodal model of post-surgical seizure-freedom outperforms any measure in isolation. Together, our results suggest that MEG band power abnormality mapping may complement current iEEG implantation strategies, providing clinically useful information to aid decision making during the pre-surgical evaluation.

Intracranial EEG recordings are used if the mapping of epileptogenic tissue using non-invasive modalities are inconclusive, discordant, uncertain of epileptogenic network involvement, or indicate a close proximity to eloquent tissue[39–41]. To minimise the risks attributed with iEEG[42], and maximise its effectiveness, a clear hypothesis of the EZ is required in order to guide electrode implantation. At present, implantation strategies are determined by clinical teams, usually based on visual evaluation of non-invasive modalities and seizure semiologies. Our MEG derived spatial maps of band power abnormalities indicate a stronger overlap between the most abnormal tissue and the implantation of iEEG electrodes in seizure-free subjects (Figure 3). As such, our data-driven abnormality maps may complement current strategies by validating the proposed electrode implantation, or by directing implantation to other brain areas.

Several studies have proposed the use of MEG recordings to help inform iEEG electrode placement[14–19,43,44]. Magnetic source imaging (MSI) indicated additional electrode coverage beyond the initially proposed hypothesis of epileptogenic tissue in 23% of subjects[16]. Moreover, in 39% of subjects, the authors report seizure-onset activity in the electrodes proposed by MSI. Frequent, and densely clustered interictal MEG spikes were correlated with iEEG placement in 69% of subjects in whom the seizure onset zone was localised[24]. Our study builds on this literature, using a data-driven framework to relate interictal MEG band power abnormalities to iEEG electrode placement without the need to mark interictal spikes.

Interictal markers of the epileptogenic zone have been developed using HFOs[7,9,10,45,46], spikes[3,4,47], and networks[48–55]. In this study we focus on the mapping of interictal band power abnormalities. Recent studies have demonstrated that the resection of the strongest abnormalities defined by iEEG[32,56] and MEG[20] in isolation are associated with post-surgical seizure-freedom. As MEG and iEEG recordings are sensitive to different types of sources[21,22] we investigated whether concordant markers of epileptogenic tissue derived from the same individuals using MEG and iEEG yielded a better resolution for the delineation of epileptogenic tissue.

Our model of post-operative seizure freedom follows an intuitive thought-process. First, MEG abnormalities must be investigated by intracranial EEG electrodes. Second, those regions must also be abnormal using iEEG. Third, those abnormal regions must be resected. If those three criteria are met then the chance of seizure freedom is extremely high. We presented our model of three properties using a nomogram, a visual tool used to illustrate complex multivariable linear models. Recent studies have proposed the use of nomograms in the context of epilepsy to aid prediction of post-operative seizure-freedom[34,35] and cognitive decline[36,37]. Our multimodal model of post-surgical seizure-freedom outperforms our single measures in isolation (AUC= 0.8). Interestingly, the feature weights of the model were roughly similar (Figure 4a), suggesting all three contribute to the best predictions of outcome. Our results indicate that MEG and iEEG band power abnormalities contain complementary information which may aid clinical decision making during the pre-surgical evaluation.

This study has several strengths and limitations. One strength is the data-driven nature of MEG and iEEG band power abnormalities, negating the need for manual spike marking, which can be prone to human bias[57]. Band power mapping however is relatively invariant to changes in spike rate and magnitude, and offers different information[32]. A key limitation of this study is the sample size, though to our knowledge is still one of the largest quantitative studies of iEEG and MEG with gold-standard postoperative MRI for resection delineation. Nontheless, future studies using larger cohorts could validate the techniques proposed. Moreover, the difficulty in localising weak signals in subcortical structures precludes the accurate analysis of individuals with seizures of temporal origin. The addition of abnormality maps derived using structural modalities such as T1 weighted MRI or diffusion MRI may circumvent the current limitation of limited coverage in deep brain structures[58].

Markers of epileptogenic tissue derived using iEEG have consistently been shown to relate to surgical outcome. Yet, iEEG implantation requires preconceived ideas of the location of epileptogenic tissue, usually acquired using qualitative techniques. We proposed interictal MEG band power abnormality mapping as a data-driven approach to complement current iEEG implantation strategies. Our findings further highlight the clinical value of MEG band power abnormalities for individuals with drug refractory neocortical epilepsy.

## Acknowledgements

We thank members of the Computational Neurology, Neuroscience & Psychiatry Lab (www.cnnp-lab.com) for discussions on the analysis and manuscript. The normative data collection was supported by an MRC UK MEG Partnership Grant, MR/K005464/1. T.O. is supported by the Centre for Doctoral Training in Cloud Computing for Big Data (EP/L015358/1). P.N.T. and Y.W. are both supported by UKRI Future Leaders Fellowships (MR/T04294X/1, MR/V026569/1). B.D. receives support from the NIH National Institute of Neurological Disorders and Stroke U01-NS090407 (Center for SUDEP Research) and Epilepsy Research UK. J.S.D is supported by the Wellcome Trust Innovation grant 218380. J.S.D and J.dT are supported by the NIHR University College London Hospitals Biomedical Research Centre, UCL Queen Square Institute of Neurology, London WC1N 3BG, United Kingdom.

# Supplementary

# Surgical outcome separability of the $D_{RS}$

The $D_{RS}$ is a measure that was recently shown to relate to post-operative seizure freedom in cohorts of individuals with refractory epilepsy[20,32]. We compute the MEG and iEEG $D_{RS}$ values for the cohort of 32 individuals with refractory neocortical epilepsy using only tissue that has coverage in both modalities. The results across the cohort, Figure S1 demonstrate the separability of surgical outcome groups based on the derived $D_{RS}$ scores. The $D_{RS}$ for both modalities performs well and in the hypothesised direction, separating surgical outcome groups with an AUC$> 0.7$.

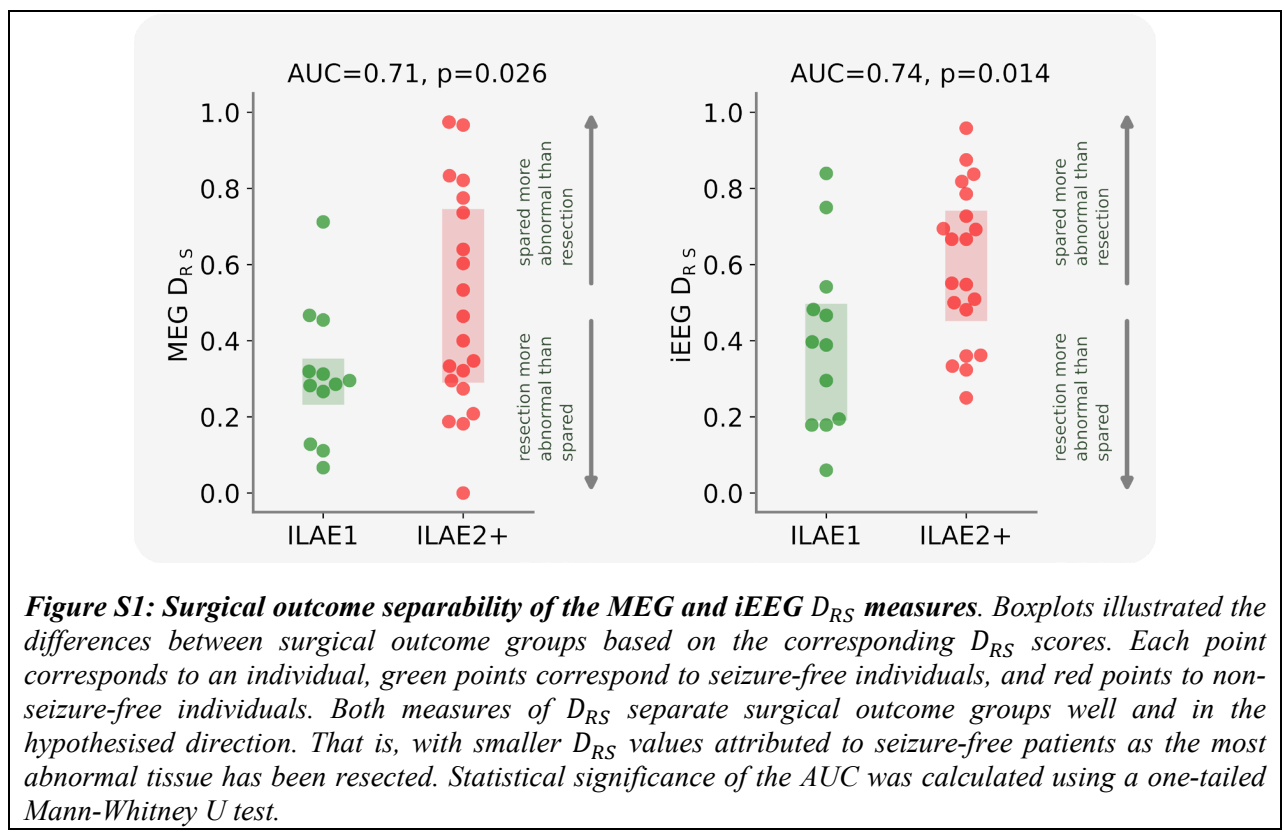

*Figure S1: Surgical outcome separability of the MEG and iEEG $D_{RS}$ measures. Boxplots illustrated the differences between surgical outcome groups based on the corresponding $D_{RS}$ scores. Each point corresponds to an individual, green points correspond to seizure-free individuals, and red points to non-seizure-free individuals. Both measures of $D_{RS}$ separate surgical outcome groups well and in the hypothesised direction. That is, with smaller $D_{RS}$ values attributed to seizure-free patients as the most abnormal tissue has been resected. Statistical significance of the AUC was calculated using a one-tailed Mann-Whitney U test.*

# Table of patient data

**Summary of patient metadata and measures.** Commonly acquired patient metadata are reported including the side of surgical resection, localisation of the resection (F: Frontal, T: Temporal, P: Parietal, O: Occipital), one year post-operative outcome. Additionally, the abnormality coverage, and both $D_{RS}$ derived using MEG and iEEG data are reported.

| Patient ID | Side | Resection Site | Surgical Outcome (1 year) | Abnormality Coverage | iEEG $D_{RS}$ | MEG $D_{RS}$ |
|---|---|---|---|---|---|---|
| 1 | L | F | ILAE 2+ | 0.299 | 0.551 | 0.000 |
| 2 | L | F | ILAE 2+ | 0.405 | 0.667 | 0.464 |
| 3 | R | F | ILAE 2+ | 0.442 | 0.548 | 0.775 |
| 4 | L | F | ILAE 1 | 0.67 | 0.482 | 0.128 |
| 5 | L | F | ILAE 1 | 0.714 | 0.840 | 0.313 |
| 6 | R | F | ILAE 2+ | 0.471 | 0.250 | 0.603 |
| 7 | R | F | ILAE 2+ | 0.716 | 0.481 | 0.188 |
| 8 | L | F | ILAE 2+ | 0.681 | 0.362 | 0.533 |
| 9 | L | F | ILAE 1 | 0.56 | 0.542 | 0.282 |
| 10 | L | F | ILAE 2+ | 0.532 | 0.694 | 0.833 |
| 11 | L | F | ILAE 2+ | 0.398 | 0.333 | 0.274 |
| 12 | R | P | ILAE 2+ | 0.608 | 0.838 | 0.347 |
| 13 | L | O | ILAE 1 | 0.476 | 0.179 | 0.455 |
| 14 | R | F | ILAE 2+ | 0.252 | 0.818 | 0.640 |
| 15 | R | F | ILAE 1 | 0.586 | 0.397 | 0.467 |
| 16 | R | P | ILAE 2+ | 0.687 | 0.324 | 0.208 |
| 17 | L | T | ILAE 1 | 0.562 | 0.389 | 0.267 |
| 18 | L | T | ILAE 2+ | 0.716 | 0.727 | 0.182 |
| 19 | L | T | ILAE 2+ | 0.537 | 0.875 | 0.400 |
| 20 | L | F | ILAE 1 | 0.63 | 0.060 | 0.295 |
| 21 | L | FP | ILAE 2+ | 0.649 | 0.500 | 0.333 |
| 22 | R | F | ILAE 2+ | 0.546 | 0.360 | 0.295 |
| 23 | L | F | ILAE 1 | 0.624 | 0.467 | 0.067 |
| 24 | L | F | ILAE 2+ | 0.508 | 0.786 | 0.821 |
| 25 | L | F | ILAE 2+ | 0.518 | 0.958 | 0.967 |
| 26 | R | OP | ILAE 1 | 0.826 | 0.194 | 0.319 |
| 27 | R | F | ILAE 1 | 0.631 | 0.295 | 0.286 |
| 28 | R | O | ILAE 1 | 0.821 | 0.179 | 0.111 |
| 29 | L | OP | ILAE 2+ | 0.649 | 0.667 | 0.321 |
| 30 | R | F | ILAE 2+ | 0.597 | 0.509 | 0.736 |
| 31 | R | P | ILAE 2+ | 0.469 | 0.692 | 0.974 |
| 32 | R | FP | ILAE 1 | 0.456 | 0.750 | 0.712 |